\documentclass[a4paper,
               biblatex,      
               keeplastbox,   
               ]{jacow}
\usepackage{pdfpages,multirow,ragged2e,graphicx}
\usepackage{caption}
\usepackage{subcaption}%
\makeatletter%
	\ifboolexpr{bool{xetex}}
	 {\renewcommand{\Gin@extensions}{.pdf,%
	                    .png,.jpg,.bmp,.pict,.tif,.psd,.mac,.sga,.tga,.gif,%
	                    .eps,.ps,%
	                    }}{}
\makeatother

\ifboolexpr{bool{xetex} or bool{luatex}} 
{}                                      
{\usepackage[utf8]{inputenc}}           

\usepackage[USenglish]{babel}

\ifboolexpr{bool{jacowbiblatex}}%
{%
  \addbibresource{bibl.bib}
 }{}
\begin{document}

\title{COMPLETION OF TESTING SERIES DOUBLE-SPOKE CAVITY CRYOMODULES FOR ESS}

\author{R.~Santiago Kern\thanks{rocio.santiago\_kern@physics.uu.se}, C.~Svanberg, K.~Fransson, K.~Gajewski, L.~Hermansson, H.~Li, T.~Lofnes, \\
A.~Miyazaki\thanks{now at CNRS/IN2P3/IJCLab, Universit\'{e} Paris-Saclay, Orsay, France}, M.~Olveg\r{a}rd, I.~Profatilova, R.~Ruber\thanks{also at Thomas Jefferson National Accelerator Facility, Newport News, Virginia, United States}, M.~Zhovner,\\ Uppsala University, Uppsala, Sweden
}
\maketitle

\begin{abstract}
The FREIA Laboratory at Uppsala University, Sweden, has completed the evaluation of 13 double-spoke cavity cryomodules for ESS. This is the first time double-spoke cavities will be deployed in a real machine. This paper summarizes testing procedures and statistics of the results and lessons learned. 
\end{abstract}

\section{THE FREIA LABORATORY}
The FREIA Laboratory in Uppsala~\cite{Ruber2021FREIA} is a leading laboratory in accelerator R\&D in Sweden and currently responsible for testing the 13 double-spoke cryomodules (plus one spare) for the European Spallation Source (ESS) in Lund~\cite{Jansson2022}. These cryomodules have two double-spoke cavities each, are assembled at Laboratoire Ir\`{e}ne Joliot-Curie (IJCLab), in Paris (France) and transported to FREIA for testing. In this regards FREIA counts with its own helium liquefaction plant and suitable radio-frequency (RF) power stations. 

\section{TESTING TIMELINE AND CAMPAIGN}

After the testing of the prototype double-spoke cryomodule finished in October 2019~\cite{Li2019RFSpokePrototype, Rocio2019CryogenicSpokePrototype,Miyazaki2020}, the infrastructure and procedures were updated based on this experience and the instrumentation adapted for the series cryomodule testing. The testing of the 13 series cryomodules started in October 2020 and has ended in May 2023, with just the spare cryomodule left to test at the time of print. From the 13 cryomodules tested, five of them had to be sent to IJCLab for repairs, bringing the total number of tests done at FREIA up to 18. Table~\ref{THIAA03_t1} gives an overview of the disqualified cryomodules and the reason for the disqualification. There were two reasons that repeated over cryomodules: 

\begin{Itemize}
\item A leak between the double wall tube and the beam vacuum at cold: due to the mechanical polishing of a weld to prepare a perfect surface for copper plating on top, and
\item A problem with the stepper motor: became mechanically non responsive after reaching certain position. The stepper motor was exchanged after warming up. The root of the cause is still under investigation by the corresponding parties. 
\end{Itemize}

The leaks in the super-critical helium (ScHe) circuit cooling the outer conductor of the coupler, which decouples the 300~K region from the 2~K, were revealed only after the cryomodules had been cooled with liquid helium. When leak tested in advance at warm there was no leak indication; thus, the cold test at FREIA was the first test identifying this issue. All of these cryomodules were tested once again at FREIA after being repaired, and all were accepted.

\begin{table}[!hbt]
   \centering
   \caption{List of disqualified cryomodules (CM) and their cause.}
   \begin{tabular}{cl}
       \toprule
       \textbf{CM \#} & \textbf{Issue}                     \\
       \midrule
           CM02         & Stepper motor lack of response        \\ 
           CM03 &     Stepper motor lack of response\\
           CM04 & Stepper motor lack of response\\
            & Vacuum leak in FPC’s double wall tube\\
            CM09 & Vacuum leak in FPC’s double wall tube\\
            CM10 & Stepper motor lack of response\\
       \bottomrule
   \end{tabular}
   \label{THIAA03_t1}
\end{table}

\subsection{Schedule}

In average, the time spent testing from arrival of the cryomodule until it is back in the cargo area is ca. six weeks, during which four of those are spent at cold. The cryomodules are shipped with the insulation vacuum and helium circuits filled with dry nitrogen gas while the cavity's beam volume is under vacuum. The cryomodule usually stays a couple of days in the FREIA hall for thermalization before any work is done. The list of tasks and the time they ideally take are, in chronological order, as follows (Fig.~\ref{THIAA03_f1}):

\begin{figure}[!htb]
   \centering
   \includegraphics*[width=1\columnwidth]{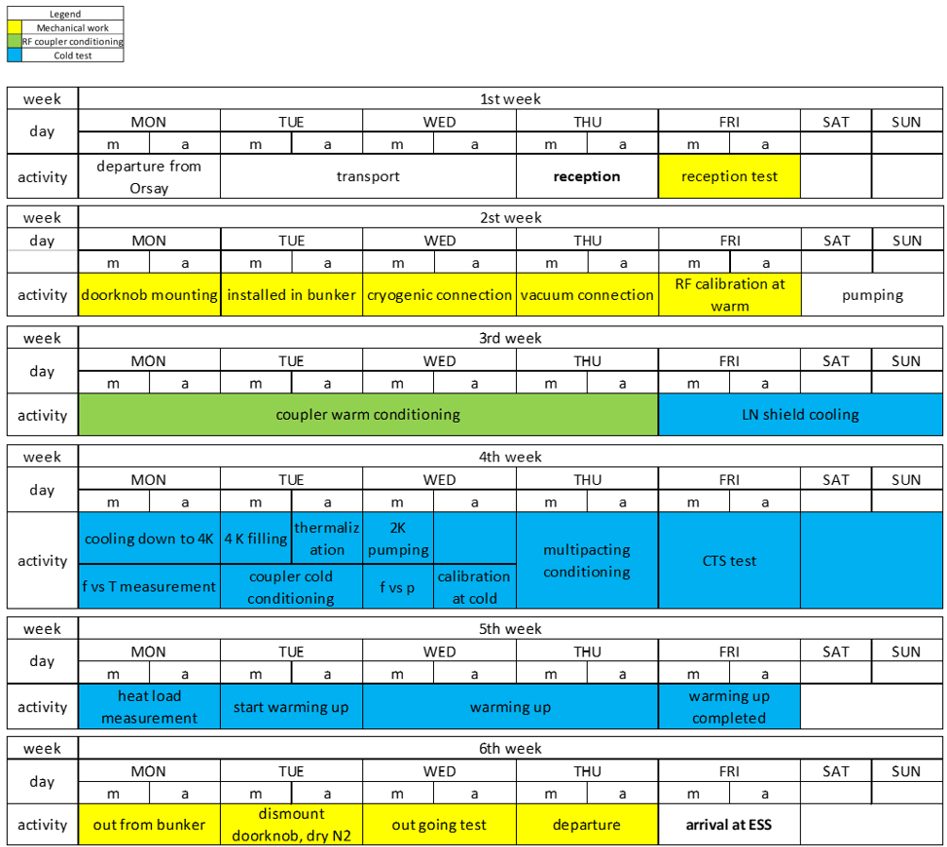}
   \caption{Testing schedule example for a cryomodule.}
   \label{THIAA03_f1}
\end{figure}

\subsubsection{In FREIA’s cargo area:}
\begin{Itemize}
\item Reception tests (1~day): to check the electrical continuity of connectors and the beam vacuum’s level. Note that the cavities arrive under vacuum (ca. $1\times10^{-3}$~mbar).
\item The cryomodule is moved to a frame with wheels and the RF waveguide-coaxial adapters (aka doorknob) are mounted (1~day).
\item The cryomodule is then wheeled into the bunker (1~day).
\end{Itemize}

\subsubsection{Once in the bunker}
\begin{Itemize}
\item The cryogenic lines of the cryomodule and the valvebox are connected (1/2~day).
\item The helium circuit is purged once to check for leaks at vacuum and above atmospheric pressure levels (1/2~day).
\item If no leaks are found, the bellow containing all cryogenic connections is closed (1/2~day).
\item The vacuum insulation for the cryomodule and the prototype valvebox is started.
\item One pumping station is connected to each cavity (1~day). This connection is done under a portable clean room whenever possible. A few days are needed to reach below $1\times10^{-7}$~mbar. 
\item The RF cables are calibrated at warm temperature (1 day).
\item Two extra purging cycles are done to clean the helium circuit from air particles.
\item The fundamental power couplers (FPCs) are RF conditioned at room temperature (4~days).
\item The cooling of the thermal shields with liquid nitrogen (LN$_2$) is started (1~day).
\item The cavities are cooled to 4.2~K with liquid helium (LHe) (1~day).
\item The FPCs are RF-conditioned at cold temperature (1~day)
\item The cavities are cooled to 2~K (1/2~day).
\item The attenuation of the field pick up line inside the cryomodule is measured and later included into the calibration. The fast interlocks for the Low-Level RF (LLRF) system are also set (1/2~day).
\item Both cold tuning systems (CTS) are tested (1~day). 
\item The RF tests of cavities are performed at 2~K (1~day). The order of RF and CTS tests is decided depending on the progress of thermalization.
\item The static and dynamic heat loads are measured with the CTS at the target position (1 day).
\item The cryomodule and the prototype valvebox are warmed up (4~days). The insulation vacuum is vented once the cavities are above 120~K to help speed up the process.
\item All cryogenic lines are disconnected (1/2~day).
\item Both beam vacuum pumping stations and RF waveguides are disconnected (1/2~day).
\item The cryomodule is wheeled out of the bunker (1/2~day).
\end{Itemize}

\subsubsection{Back in FREIA’s cargo area:}
\begin{Itemize}
    \item The doorknobs are dismounted (1~day).
    \item The insulation vacuum is filled with gas nitrogen.
    \item Departure tests (1/2~day): to check the electrical continuity of connectors and the beam vacuum’s level before shipment. 
    \item Load cryomodule in box (½~day).
    \item Shipment to ESS or IJCLab with shock loggers mounted on FPCs.
\end{Itemize}

More detailed information on these tasks is given in the following sections.

\section{HANDLING AND INSTALLATION}
After the reception tests, the cryomodule is placed on a support frame equipped with wheels to be able to move it around inside the lab. The doorknobs are already put in place on the support frame before the cryomodule, but in a lowered position to not clash with the ceramic windows. The doorknobs are then carefully lifted so that the outer conductor can be mounted to the ceramic window flange. With the lower flange of the outer conductor not connected it is possible to rotate the doorknobs to align them. After alignment, all flanges are tightened and the supporting structure locked to prevent any stress on the ceramic windows. The inner conductor is then pushed up and tightened together with the pipe for cooling water\footnote{The doorknobs used for tests at FREIA are prototypes and different from the series ones with a potential DC-bias option in the tunnel at ESS.}. A leak check is performed by connecting a hose to the cooling circuit of the antenna and flowing water through it.

To speed up the preparation work two support frames are used at FREIA, so that when a cryomodule is tested in the bunker it is possible to dismount and ship the previous one and then receive and prepare the next cryomodule for test. When a cryomodule test is finished inside the bunker it is moved to the docking area where the next prepared cryomodule is waiting. Testing equipment like turbo pumps, safety valves and pressure gauges are moved from the tested cryomodule to the prepared cryomodule while transport locks and flanges are moved from the prepared cryomodule to the already tested one.

When a cryomodule is installed in the bunker the doorknobs are first aligned and connected to the fixed waveguides coming out from underneath the prototype valvebox. Then the position of this valvebox can be adjusted so that the cryogenic lines match. A leak check of the helium cryogenic connections, consisting of three steps, is performed. First step is to purge and fill the helium circuit with clean helium gas up to around 1150~mbar. Step two involves sniffing the connections with the leak detector. For the last step the helium circuit is left with the 1150~mbar over-pressure, usually overnight, to see if the pressure changes over time. If no leaks are found and pressure is stable the multilayer insulation material and thermal screen are installed and the bellow connecting the cryomodule's and valvebox's insulation vacuum closed. 

\section{Fundamental Power Coupler Conditioning}
The FPCs, pre-conditioned at IJCLab in pairs via travelling wave up to 400~kW and standing wave up to 170~kW in a dedicated coupler conditioning bench, are also conditioned in-situ inside cryomodules at FREIA with a standing wave up to 400~kW at both room temperature and 2~K~\cite{Miyazaki2020}.

\begin{figure}[!htb]
   \centering
   \includegraphics*[width=.75\columnwidth]{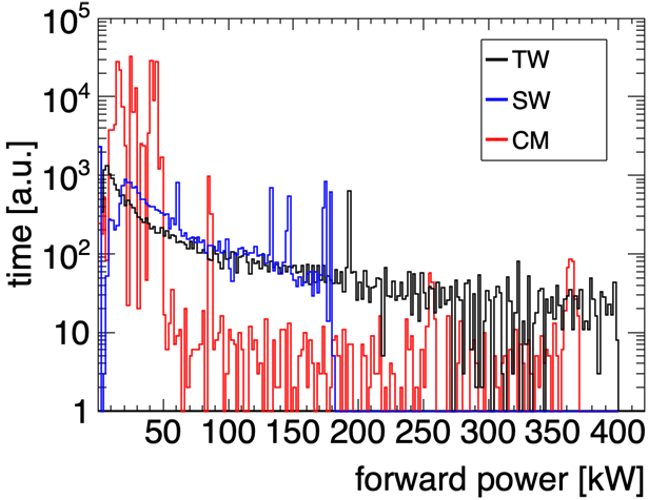}
   \caption{Comparison of the FPC conditioning back-to-back with standing and travelling wave (IJCLab) and when assembled in the cryomodule (FREIA).}
   \label{THIAA03_f2}
\end{figure}

Fig.~\ref{THIAA03_f2} shows a typical comparison of all three conditioning processes for the FPCs of two cavities and then when placed in their corresponding cryomodule 
When the FPCs are conditioned in the cryomodule, the outgassing is mainly below 100~kW and the travelling and standing wave conditioning have a different field distribution from in-situ conditioning inside the cryomodule, so that the ceramic window in the FPC is conditioned before being assembled to the cavities.

\begin{figure}[!htb]
   \centering
   \includegraphics*[width=1\columnwidth]{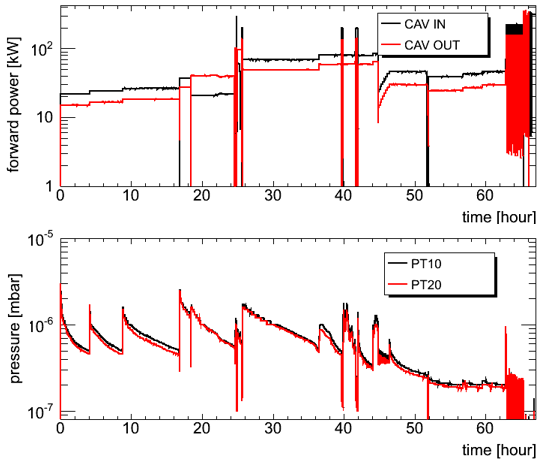}
   \caption{Details of the applied RF power and corresponding pressure during the conditioning of the FPCs (from CM05).}
   \label{THIAA03_f3}
\end{figure}

Fig.~\ref{THIAA03_f3} shows an example of the FPC conditioning for each cavity in one of the cryomodules at FREIA. 
At the starting point a short pulse length (50~$\mu$s) is chosen and the power increased until a defined first vacuum threshold (typically $5\times10^{-7}$~mbar) is reached, the power is then maintained. 
If the vacuum continues to deteriorate and goes above a defined second threshold (typically $1\times10^{-6}$~mbar) the power is reduced by half, and if the vacuum is not restored to a certain level the power is cut off. 
The power is gradually increased up to 320~kW, or full power (ca. 400~kW) if needed. 
Once the conditioning at this pulse length has been completed, this parameter is increased and the process started again~\cite{Li2021CMStatistic}. The whole conditioning process will finish when the pulse length reaches 3.2~ms (duty cycle of 4.5\%). We typically perform several extra cycles at this maximum pulse length to thoroughly get rid of the outgassing. The repetition rate is fixed at 14~Hz. It can take between three to four full days to completely clean the outgassing until reaching a duty cycle of 4.5\%. This also means that the RF stations must have high reliability and stability to avoid delays.

Due to the problems found when testing the first time, as described in Table~\ref{THIAA03_t2}, some power couplers have been conditioned twice: some of them without being exposed to air in between (stepper motor issues) and some other power couplers, since they were repaired and recycled from disqualified cavity strings due to a vacuum leak (Table~\ref{THIAA03_t1}), were vented, re-preconditioned and re-fitted to a new cavity. The reason for CM02 taking up to 112~hours was due to the beam vacuum not being evacuated for quite some time (the valve to the pumping station was closed). In the case of CM03, it took 109 h because there was only one pumping station connected and for CM04, only one RF station was available at the time. Except for these three exceptional cases, the conditioning time ranges from just 10~h to up to 100~h.

\begin{table}[!hbt]
   \centering
   \caption{List of power FPCs used in each cryomodule tested in chronological order, with the number of pumping stations used and the total conditioning hours. Marked in red are the FPCs that were repaired and retested.}
   \begin{tabular}{lcccr}
       \toprule
       \textbf{CM \#} & \textbf{FPC1} & \textbf{FPC2} & \textbf{\# Pumps} & \textbf{Hours}                     \\
       \midrule
          
CM02	& CPL01&	CPL04 &	1 &	112\\
CM04 & 	\textcolor{red}{CPL11} &	\textcolor{red}{CPL03} &	1 &	67\\
CM02 (2nd) &	CPL01&	CPL04&	2	&9\\
CM05&	CPL14	&CPL18	&2	&66\\
CM03	&CPL06	&CPL26	&1	&109\\
CM01	&CPL10	&CPL12	&2	&90\\
CM04 (2nd)	&CPL32	&CPL05	&2	&147\\
CM03 (2nd)	&CPL06	&CPL26	&2	&12\\
CM06	&\textcolor{red}{CPL11}	&CPL20	&2	&66\\
CM07	&CPL25	&CPL30&	2	&48\\
CM08	&CPL21	&CPL15	&2	&65\\
CM09	&\textcolor{red}{CPL27}	& \textcolor{red}{CPL28}	&2	&30\\
CM10	&CPL23	&CPL24	&2	&10\\
CM11	&CPL22	&CPL19	&2	&26\\
CM12	&\textcolor{red}{CPL03}	&CPL09	&2	&92\\
CM10 (2nd)	&CPL23	&CPL24	&2	&9\\
CM09 (2nd)	&CPL16	&CPL17	&2	&67\\
CM13	&\textcolor{red}{CPL27}	& \textcolor{red}{CPL28}	&2	&100\\
       \bottomrule
   \end{tabular}
   \label{THIAA03_t2}
\end{table}

\begin{figure}[!htb]
   \centering
   \includegraphics*[width=1\columnwidth]{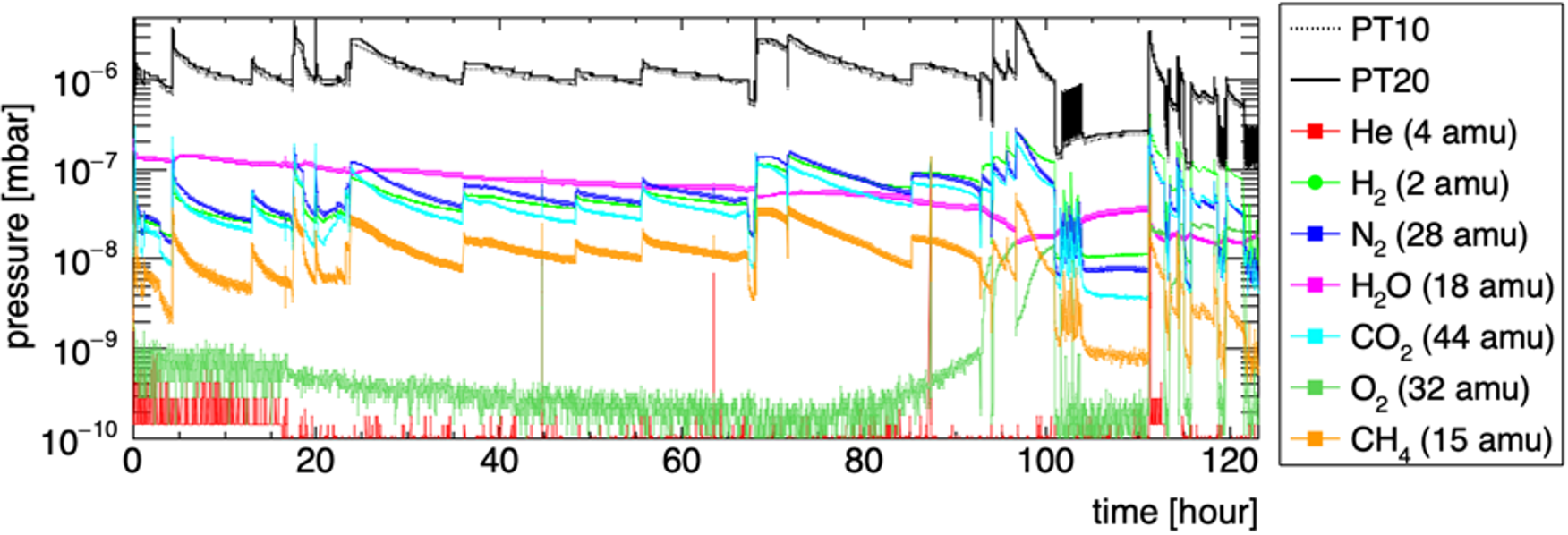}
   \caption{The history of gas species decomposed by the residual gas analyzer (RGA) during the FPCs conditioning.}
   \label{THIAA03_f4}
\end{figure}

As a side note, during the conditioning a residual gas analyzer (RGA) is connected to one of the pumping stations to mainly trace potential helium gas leaks, but sometimes oxygen and other species, including hydrocarbons, have been observed to have a negative correlation which might hint at a chemical reaction between them (Fig.~\ref{THIAA03_f4}). 
Its cause and effect in the FPCs conditioning process are still unclear.

\section{CRYOMODULE COOLING}

\begin{figure}[!htb]
   \centering
   \includegraphics*[width=1\columnwidth]{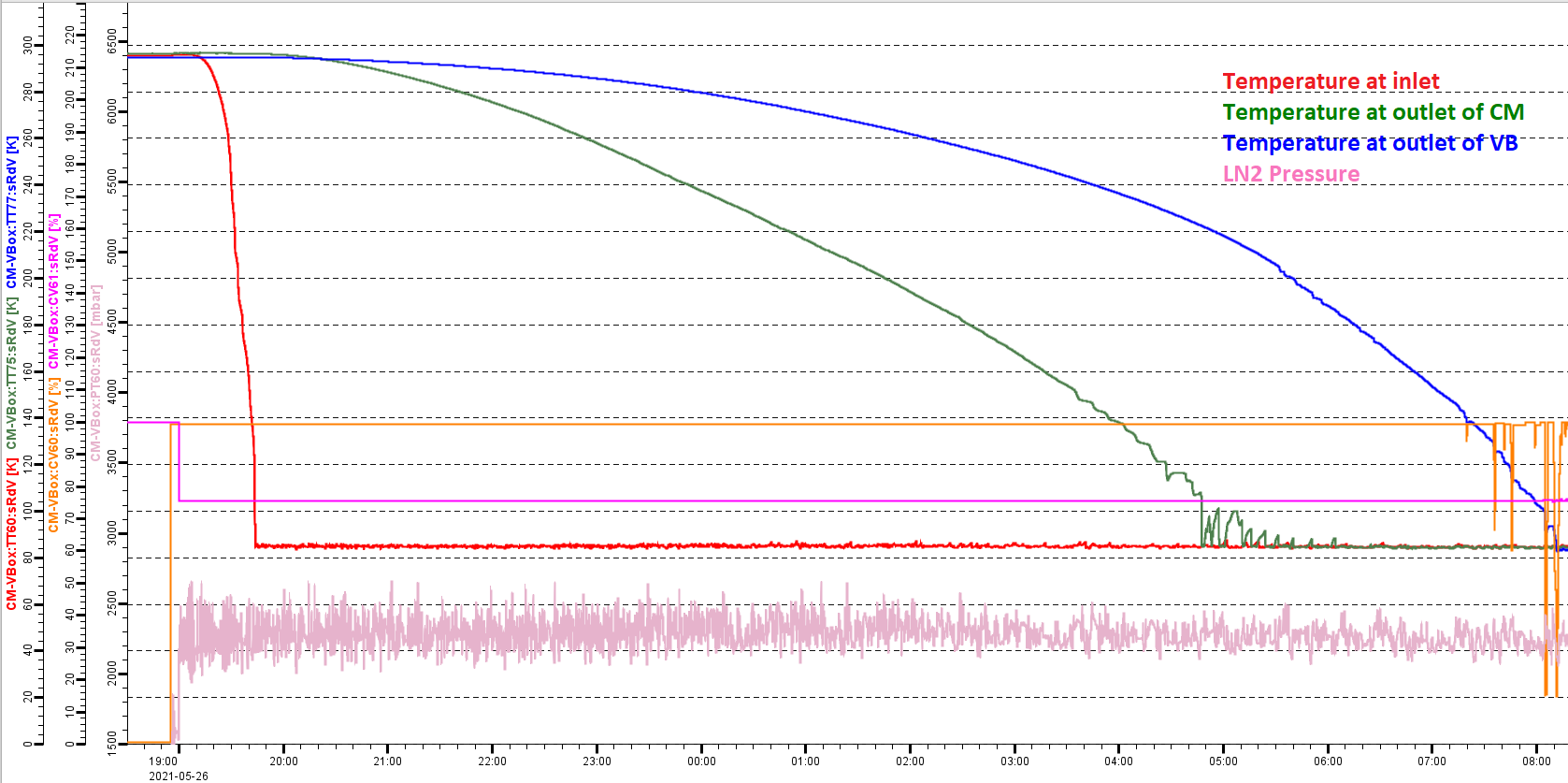}
   \caption{Cooldown details of the thermal shield for the prototype valvebox and cryomodule with liquid nitrogen.}
   \label{THIAA03_f5}
\end{figure}

\begin{figure}[!htb]
   \centering
   \includegraphics*[width=1\columnwidth]{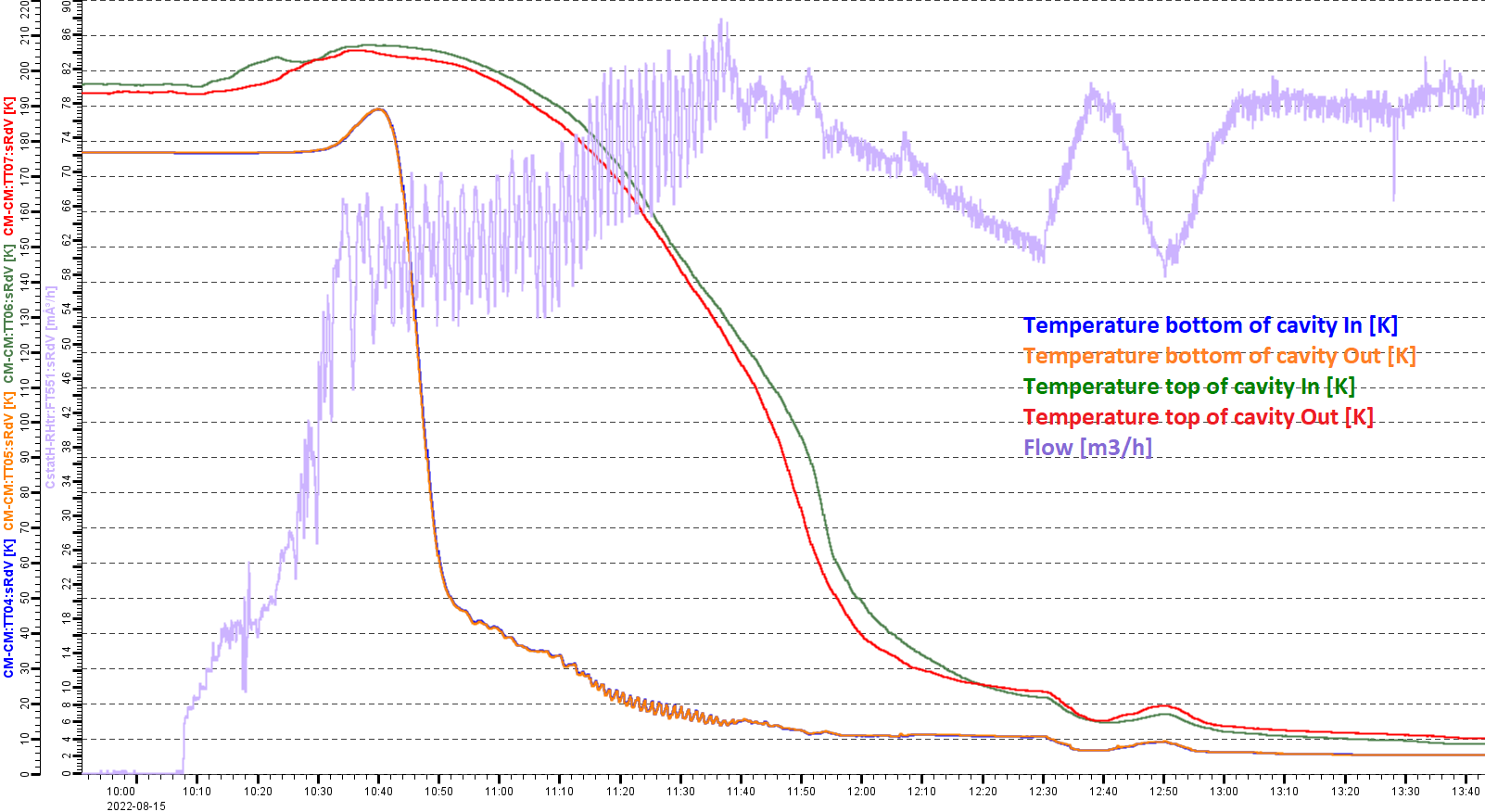}
   \caption{Liquid helium cooldown details for both cavities (CM11).}
   \label{THIAA03_f6}
\end{figure}

As mentioned before, LN$_2$ is used for cooling the cryomodule’s and prototype valvebox’s thermal shield to about 80~K. Note that the thermal shields consist of only one cooling circuit for both, entering and exiting through the prototype valvebox. Once the shields are cold, usually 12~h later (Fig.~\ref{THIAA03_f5}), the cooling with LHe at 4.2~K can start (Fig.~\ref{THIAA03_f6}), but if done back to back the thermalization of the system is still ongoing and it can take up to 2~days before there is LHe level stabilized at the target volume in the helium tank. After the cavities are filled with LHe at 4.2~K, the conditioning of the FPCs can start. Afterwards, the temperature is lowered to 2~K by reducing the pressure in the gas return line to 31~mbar via several vacuum pumps (labelled sub-atmospheric) located 20~m away from the cryomodule. Still several more days are needed for the thermalization of the CTSs, when the temperature at the motor side is around 100~K. It is at this point that the CTS tests and the heat load measurements are done.  

\subsection{Cavities Cooldown Rates}

\begin{figure}[!htb]
   \centering
   \includegraphics*[width=.8\columnwidth]{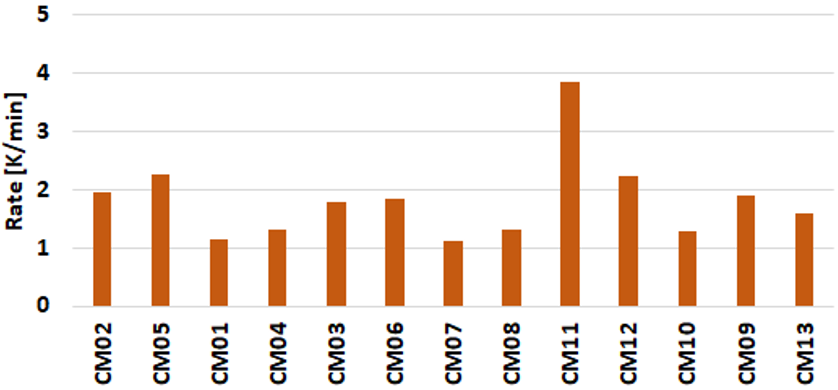}
   \caption{Cooldown rate for the successfully tested cryomodules, in chronological order.}
   \label{THIAA03_f7}
\end{figure}

One parameter that was considered statistically interesting was the cooldown rate of the cavities, measured between 150~K and 50~K. This value varies greatly among cryomodules, as shown in Fig.~\ref{THIAA03_f7}, where only the data for the cryomodules qualified after this particular cooling down is shown in chronological order. There are several factors affecting the cooling rate, such as pre-thermalization by LN$_2$. In this case, whenever possible modules were thermalized for longer times than others by profiting from a weekend, but not all. Factors affecting external equipment related to the testing were a degradation of the circulation compressor for unknown reasons (a bottleneck for some of the modules) and the low capacity of the recovery compressors. During the series testing the recovery compressor system was equipped with an extra compressor with triple the capacity. This upgrade was in used from CM11 onwards. Although such uncontrolled cooling down rates have a wide spread, this did not reveal any individual difference among modules and posed no problem for normal operation.

\section{Static and Dynamic Heat Loads}
The static and dynamic heat loads in the 2~K operation are measured via a volumetric flowmeter at room temperature and atmospheric pressure, located after the sub-atmospheric pumps. 
During this measurement the filling valve to the cavities is closed, and due to the construction of the prototype valvebox at FREIA, the cooling flow towards the FPCs is reduced, increasing its temperature and thus the heat load onto the cavity. Instead of a dedicated super critical helium (ScHe) line, which is lacking at this moment, the FPCs cooling was achieved by rerouting part of the return gas helium from the 4~K tank (located in the prototype valvebox), which is at 4.2~K and 1.1~bara. 
Moreover, as already mentioned, the thermal shield is cooled by LN$_2$ in FREIA, thus the measured heat load can differ from the future operation in ESS where ScHe will be used instead and the shield temperature will be around 40~K.

Unfortunately, not all cryomodules have been measured under the same conditions due to different issues, the main one being the presence of thermo-acoustic oscillations (TAOs). These oscillations seem to be related to FREIA's specific system configuration (prototype valvebox, LN$_2$ cooling, lack of ScHe circuit) and does not imply an intrinsic issue in module design and its operation in the ESS machine. These oscillations could be reduced, but not damped, by reducing the opening of some of the valves in the return gas line. This change of valve opening among cryomodules led to an impact in the total flow measured, but it had no effect in the difference between static and dynamic. Another change done to the system, and also related to TAOs, was the valve used for regulating the pressure at 2~K: it was changed in favour of a valve closer to the cavities.

\begin{figure}[!htb]
   \centering
   \includegraphics*[width=.8\columnwidth]{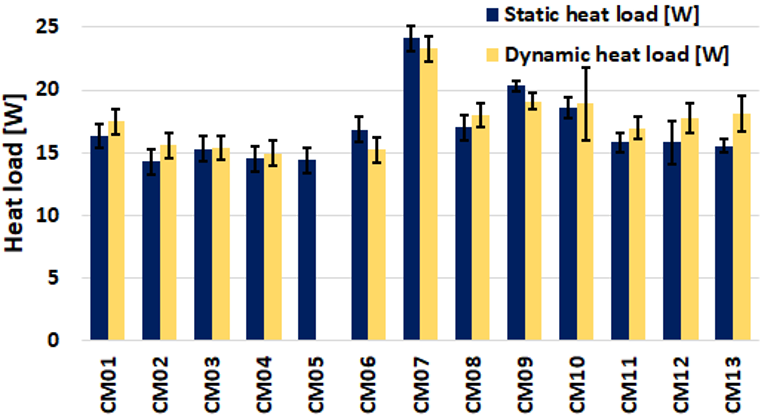}
   \caption{Static and dynamic (9 MV/m for both cavities) heat load statistics for the accepted cryomodules.}
   \label{THIAA03_f8}
\end{figure}

Fig.~\ref{THIAA03_f8} gives the final values for the static and dynamic heat loads with both cavities at an accelerating gradient of 9~MV/m for the accepted cryomodules. There is no dynamic heat load value for CM05 since the cavities could not be powered at the same time (one of the power stations was in maintenance). Although the static and dynamic heat loads are within the error bars, the reason for having at times the dynamic value lower than the corresponding static might come from the system not being sufficiently thermalized at 2~K. Depending on the schedule, the static heat load has been sometimes measured a couple of days before the cavities were powered. The high error measured in CM10 is due to field emission onset for both cavities at the nominal field. In any case, from this figure it can be seen that, in overall, the difference between the dynamic and static heat loads is zero consistent within the error and, therefore, the RF power loss of the cavities is maximum 2~W in the most conservative upper limit.

\begin{figure}[!htb]
   \centering
   \includegraphics*[width=.7\columnwidth]{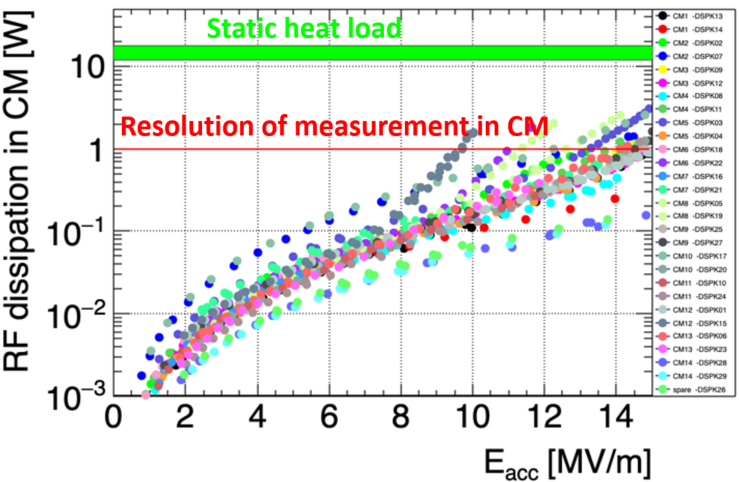}
   \caption{Power dissipated in the cryomodules at different accelerating gradients. An average of the static heat load and the measurement resolution are added for comparison purposes.}
   \label{THIAA03_f9}
\end{figure}

Since dynamic heat load is the dissipation of RF power on cavities' walls, the cause of power dissipation is the sum of non-zero surface resistivity of a superconducting material plus parasitic effects due to imperfections in the surface preparation of the cavity, such as field emission and multipacting. 
Not observing significant RF power dissipation from the cavities at the nominal gradient seems to be consistent to the prediction from the vertical tests at IJCLab (Fig.~\ref{THIAA03_f9}). 
Since the unloaded quality factor Q$_0$ is one order of magnitude higher than the specification and the duty cycle is only 4.5\%, the expected RF heat loss is of the order of mW (after multipacting conditioning), which is much lower than the measurement resolution of the calorimetric method at FREIA. 
This indeed proves that cavity performance has not been significantly degraded after the vertical tests.

\section{FIELD EMISSION}

\begin{figure}[!htb]
   \centering
   \includegraphics*[width=1\columnwidth]{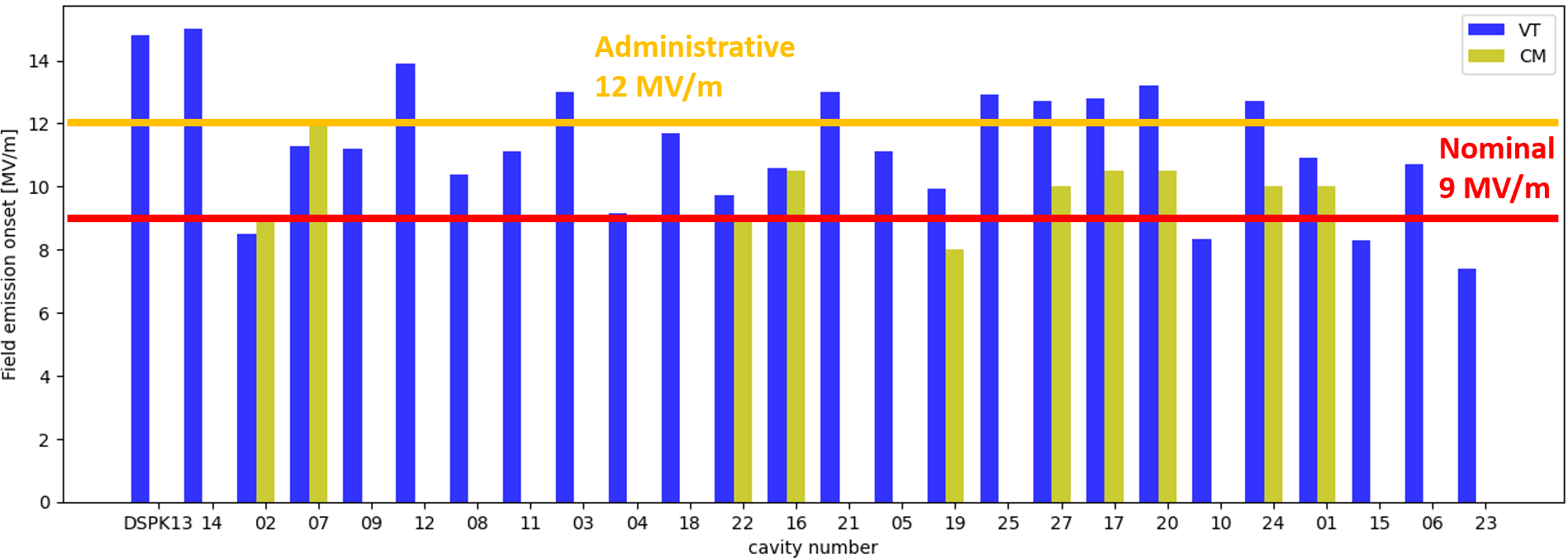}
   \caption{Field emission onset cavity (DSPK\#) comparison between vertical VT (IJCLab) and cryomodule CM (FREIA) tests.}
   \label{THIAA03_f10}
\end{figure}

Considering the individual cavities, an important parameter is the field emission onset. After the vertical tests at IJCLab, the cavities did not go through a high pressure water rinsing (HPR) process before being assembled in the cryomodules. A comparison between the field emission onset measured per cavity at IJCLab during the vertical tests~\cite{Miyazaki2023} and measured at FREIA in the cryomodules is given in Fig.~\ref{THIAA03_f10}. Here the cavities are individually named and every pair is assembled into one cryomodule, starting at CM01 and ending in CM13. The reason for not seeing strong field emission for cavity number 23 once assembled in the cryomodule even though it showed field emission at a low accelerating gradient in the vertical test, is due to the fact that, as an exception, this cavity (together with number~6) was successfully cleaned after the vertical test but could not be measured again in the vertical cryostat afterwards. 

Since most of the cavities show no clear field emission up to 12~MV/m (30\% safety margin) when tested in the corresponding cryomodule, and all cavities reached the nominal value of 9~MV/m, the strategy chosen by IJCLab together with a tight quality control of venting, demounting, and assembly of the cavities in a clean room proves to be successful.

\begin{figure}[!htb]
   \centering
   \includegraphics*[width=.7\columnwidth]{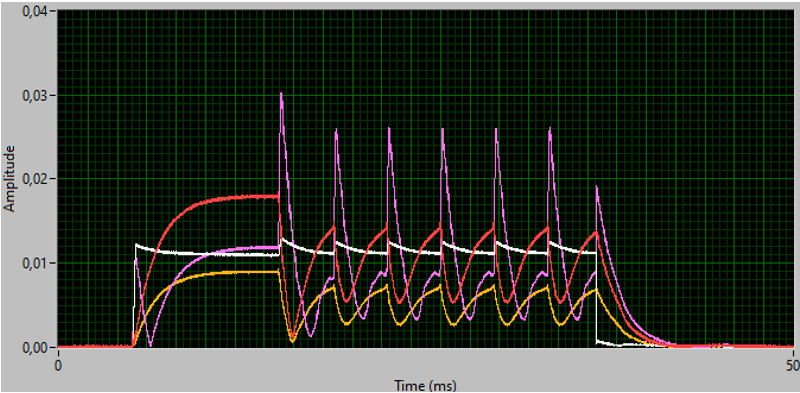}
   \caption{RF pulse shape distortion by repetitive local thermal quenching (in white - forward, red - transmitted, violet - reflected power).}
   \label{THIAA03_f11}
\end{figure}

\section{Quench detection}
When conditioning the cavities there is always a risk of a quench, even at lower voltages, which is found to be a unique feature and risk of spoke cavities compared to conventional elliptical cavities.
Significant distortions of RF pulse shape were observed at very low fields (<1~MV/m), see Fig.~\ref{THIAA03_f11}. This behaviour can be explained by local thermal quenching, perhaps induced by strong multipacting. Local thermal quench can easily become global and quench a cavity without careful low power conditioning and appropriate interlock protection. For this reason, a system that tries to determine if a quench is approaching has been developed at FREIA. This is done by calculating the decay time of the cavity field at the end of every cycle after the RF signal is turned off~\cite{Branlard2013}. If the decay time is shorter than two thirds of the measured decay time, the likelihood of a quench is considered too high and the control system will enable an interlock to interrupt the RF signal to the cavity. The calculation of the decay time is done in less than 20~\(\mu\)s in an FPGA. And if the result of the calculation at the end of a cycle is that an interrupt is needed, the RF is turned off before the start of the next cycle.

\begin{figure}[!htb]
   \centering
   \includegraphics*[width=.4\columnwidth]{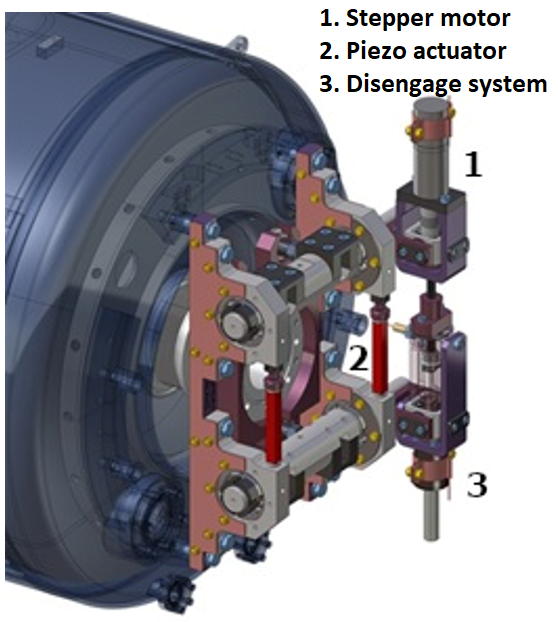}
   \caption{CAD model of the CTS (courtesy IJCLab).}
   \label{THIAA03_f12}
\end{figure}

\section{Cold Tuning System}
\begin{figure}[!htb]
   \centering
   \includegraphics*[width=1\columnwidth]{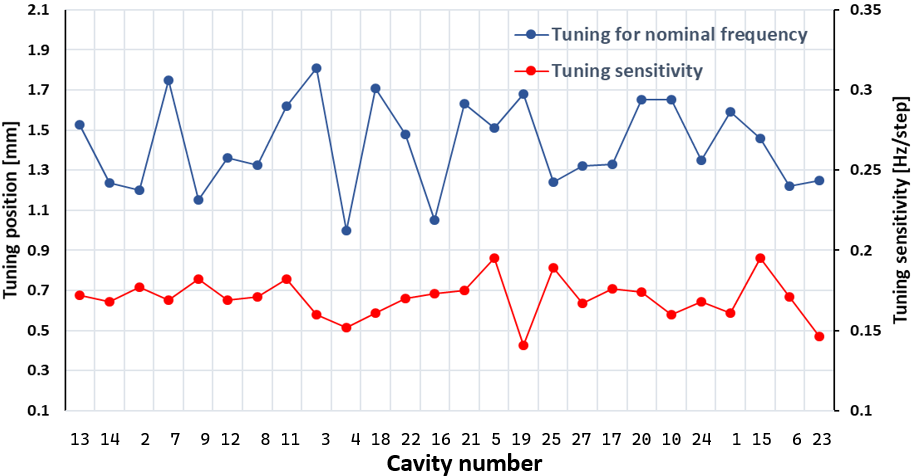}
   \caption{Motor tuning distance to nominal frequency and sensitivity.}
   \label{THIAA03_f13}
\end{figure}

The cold tuning system (CTS) consists of a stepper motor actuator for slow tuning and two piezo actuators for fast tuning~\cite{Gandolfo2013DeformationCavity,Gandolfo2014FastTunner}, see Fig~\ref{THIAA03_f12}. 
The slow tuners, used to tune the cavity frequency to its nominal value, were tested by moving forward to the target frequency and then back again in a series of steps.
For the 26 cavities measured (with four stepper motors replaced), Fig.~\ref{THIAA03_f13} shows the movement of the tuner (in mm) required to reach the target 352.21~MHz, affected by the starting frequency and the stiffness of the cavity, and the cavity's sensitivity. 
All cavities reached the target within the specified limits, and also the function of the limit switch, which gives a reference for the starting position of the motor, was tested both at warm and 2~K.

The two fast linear piezo actuators are needed to actively compensate for nonlinear behaviour of the cavity during the RF pulse, especially Lorentz force detuning (LFD). The RF electromagnetic field inside the cavity results in a radiation pressure on the cavity walls. This force deforms the cavity and shifts the resonant RF frequency of the cavity. This frequency shift, or detuning, can be calculated from the measurements of the power sent to the cavity and the measurements of the field inside. 

\begin{figure}[!htb]
   \centering
   \includegraphics*[width=.7\columnwidth]{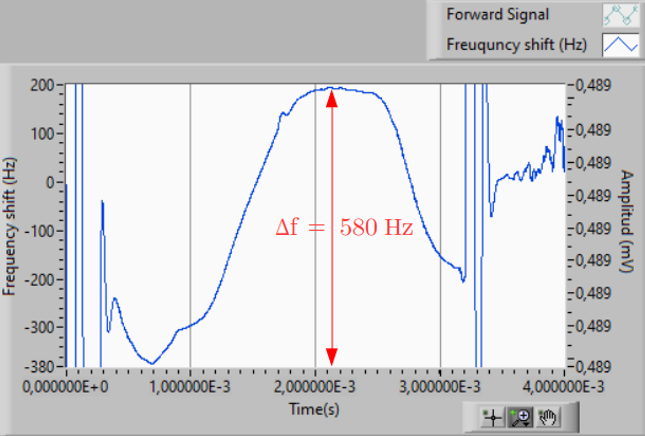}
   \caption{Frequency shift due to Lorentz force detuning. \(\Delta\)f is the difference between the maximum frequency and the minimum frequency during a cycle.}
   \label{THIAA03_f14}
\end{figure}

\begin{figure}[!htb]
   \centering
   \includegraphics*[width=.8\columnwidth]{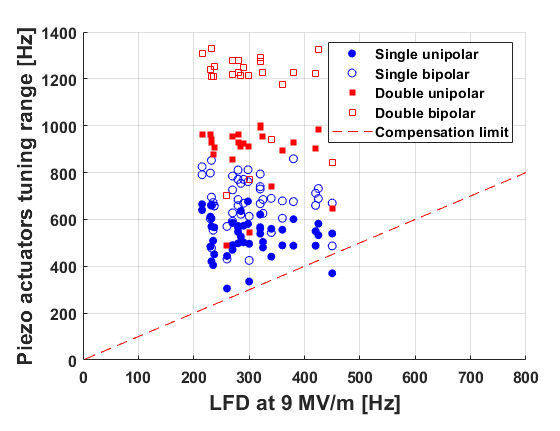}
   \caption{Measured tuning range of the piezo actuators in Hz with respect to hosting cavity's LFD at 9 MV/m.}
   \label{THIAA03_f15}
\end{figure}

Of interest is the difference between the maximum and the minimum frequency, \(\Delta\)f, during a cycle, see Fig.~\ref{THIAA03_f14}. This difference in frequency is compared to the frequency range of the piezo tuner, where the range of the tuner needs to be larger than \(\Delta\)f. The piezo actuators can be operated in unipolar (0~V to +200~V) or bipolar mode (-40~V to +200~V). A static test for each piezo actuator was performed during series testing, with the outcome shown in Fig.~\ref{THIAA03_f15}. From this plot it is clear that a tuning value for a single piezo actuator is more than 400~Hz, which is 20\% higher than the typical LFD at 9~MV/m, fully covering the needed range during pulse operation at the nominal accelerating field.

\section{LESSONS LEARNED}

After testing series cryomodules for two years and a half, FREIA has now acquired great competence and expertise in such a wide field, but it has not come for free. 
One of the points that were addressed and executed mid testing was the low capacity of the high pressure compressors. Although not a bottleneck for this series testing, it helped reducing the time spent cooling down. 
With regards to heat load measurement, and currently under discussion at FREIA, is the placement of a new flowmeter with a lower range in parallel with the existing one, which might prove very useful when measuring heat loads in the future. 
In terms of power, FREIA used two 400~kW tetrode base amplifiers, designed as proof of concept for future ones at ESS. 
Reaching stable long term operation was challenging: the number of failures with tetrodes, power supplies and amplifiers cause difficulties to follow a strict schedule for series cryomodule testing~\cite{Zhovner2021StatusCryomodules,Li2021CMStatistic,Maiano2021LINAC2022}. 
Some motors in the CTS did not work well in a few series cryomodules.
Because of the important role these motors play together with a cavity in an accelerator, it is of utmost importance to stress test motor prototypes as well as, at least, one series motor out of all the production batch. 
This is an important feedback to cryomodule manufacturers.

Finally, careful optimisation of operation parameters, risk analysis and availability of spare parts is necessary to successfully finish the testing. 
In day-to-day activities, this testing has made it obvious that a good planning is essential, and a good understanding of what processes and activities can be done or prepared in parallel will help greatly if time is of importance. 

\section{CONCLUSIONS}
FREIA successfully completed the site acceptance tests of the series double-spoke cavity cryomodules for ESS.
The measurements include mechanical, vacuum, electric tests and calibration, conditioning outgassing from FPCs, thermal shield cooling with LN$_2$, liquid helium cooling and pumping down to 31~mbar, tuner tests, RF tests of cavities and heat load estimations.
Coordination of work and the overall procedure, flexibly updated over time, were key for an efficient continuation of the series tests.
Although a few cryomodules were disqualified in the first tests,
all these modules passed the second test after repairing the faulty components.
Also, no major performance degradation of cavities was observed compared to the vertical tests before cryomodule assembly.
Our tests revealed partially unexpected challenges specific to spoke cavities in series cryomodules, such as multipacting and importance of fast interlock even at a very low field, and our experiences will be invaluable feedback to future spoke cavity projects in the SRF community.

\section{acknowledgments}
We are grategul to N.~Gandolfo and G.~Olry for the fruitful discussions and also would like to acknowledge with appreciation the crucial role of our colleagues from ESS. Special thanks go to all the technical and administrative staff at Uppsala University for their great cooperation in conducting the ESS spoke-cavity cryomodule project.

\ifboolexpr{bool{jacowbiblatex}}%
	{\printbibliography}%
	{%
	

} 
%
%


\end{document}